\documentclass[10pt]{article} 
\usepackage[preprint]{tmlr}

\usepackage{amsmath,amsfonts,bm}

\def\eqref#1{equation~\ref{#1}}

\def\1{\bm{1}}

\DeclareMathAlphabet{\mathsfit}{\encodingdefault}{\sfdefault}{m}{sl}
\SetMathAlphabet{\mathsfit}{bold}{\encodingdefault}{\sfdefault}{bx}{n}

\usepackage{wrapfig}
\usepackage{graphicx} 
\usepackage[utf8]{inputenc} 
\usepackage[T1]{fontenc}    
\usepackage{hyperref}       
\usepackage{url}            
\usepackage{booktabs}       
\usepackage{amsfonts}       
\usepackage{nicefrac}       
\usepackage{microtype}      
\usepackage{xcolor}         
\usepackage{amssymb}

\usepackage{amsthm}
\usepackage{amsmath} 

\theoremstyle{definition}
\newtheorem{dfn}{Definition}

\newtheorem{thm}{Theorem}

\title{Uncertainties in Physics-informed Inverse Problems: The Hidden Risk in Scientific AI}

\author{\name Yoh-ichi Mototake \email y.mototake@r.hit-u.ac.jp \\
      \addr Graduate School of Social and Data Sciences\\
      Hitotsubashi University\\
      Tokyo, 186-8601, Japan
      \AND
      \name Makoto Sasaki \email sasaki.makoto@nihon-u.ac.jp \\
      \addr Department of Electrical and Electronic Engineering\\
      Nihon University\\
      Chiba, 274-0072, Japan}

\def\month{MM}  
\def\year{YYYY} 
\def\openreview{\url{https://openreview.net/forum?id=XXXX}} 

\begin{document}

\maketitle

\begin{abstract}
Physics-informed machine learning (PIML) integrates partial differential equations (PDEs) into machine learning models to solve inverse problems, such as estimating coefficient functions (e.g., the Hamiltonian function) that characterize physical systems. 
This framework enables data-driven understanding and prediction of complex physical phenomena. 
While coefficient functions in PIML are typically estimated on the basis of predictive performance, physics as a discipline does not rely solely on prediction accuracy to evaluate models. 
For example, Kepler’s heliocentric model was favored owing to small discrepancies in planetary motion, despite its similar predictive accuracy to the geocentric model. 
This highlights the inherent uncertainties in data-driven model inference and the scientific importance of selecting physically meaningful solutions. 
In this paper, we propose a framework to quantify and analyze such uncertainties in the estimation of coefficient functions in PIML. 
We apply our framework to reduced model of magnetohydrodynamics and our framework shows that there are uncertainties, and unique identification is possible with geometric constraints. 
Finally, we confirm that we can estimate the reduced model uniquely by incorporating these constraints.
\end{abstract}

\section{Introduction}
\begin{figure}[bth]
\begin{center}
 \includegraphics[width=0.6\textwidth]{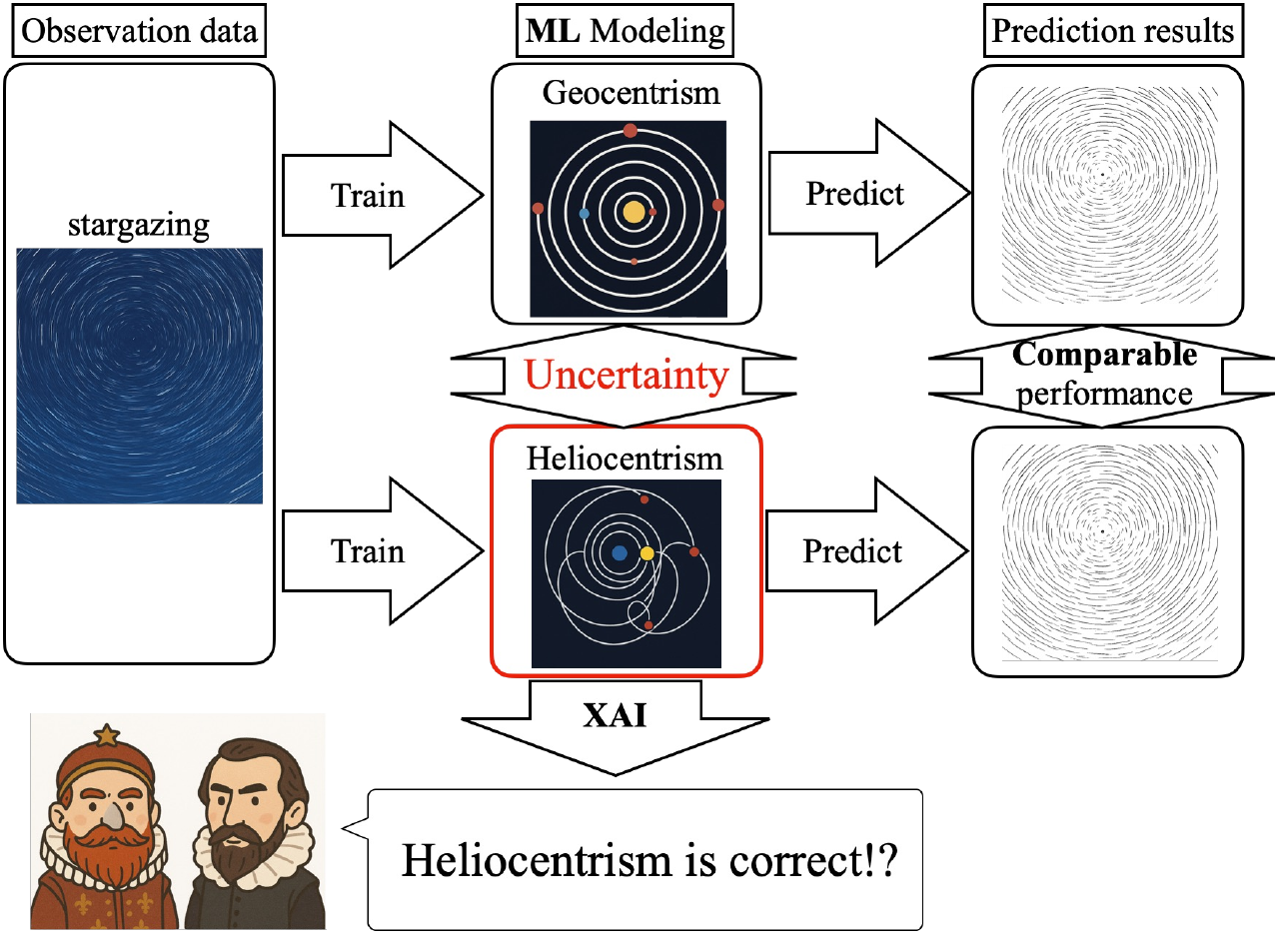}
\end{center}
\caption{Risks of scientific research using machine learning. In the presence of an uncertainty, a machine learning model may sometimes provide an interpretation that is physically unfavorable. 
(The figure was generated using DALL-E3, OpenAI)\label{fig1}}
\end{figure}

There is active research attempting to elucidate the laws of physics in a data-driven manner using machine learning methods~\citep{karniadakis2021physics,hao2022physics}. 
If the basis functions of the physical laws are known to some extent, it has been reported that physical laws can be extracted from time series data of dynamical systems by using linear regression models~\citep{brunton2016discovering} or symbolic regression methods~\citep{udrescu2020ai}. 
Even when there is limited a priori information, such as basis functions, research is being conducted to combine models with high expressive power, such as deep neural networks (DNNs), with its interpretation~\citep{arrieta2020explainable, love2023explainable} to give interaction laws~\citep{cranmer2020discovering} or conservation laws~\citep{kaiser2018discovering, wetzel2020discovering, liu2021machine, ha2021discovering, PhysRevLett.128.180201, mototake2021interpretable, PhysRevLett.128.180201, zhang2021learning, lu2023discovering, mototake2021interpretable} for the system. 
Whereases the parameters of machine learning models are typically estimated on the basis of predictive performance, physics as a discipline does not rely solely on prediction accuracy to evaluate models. 
For example, Kepler’s heliocentric model was favored owing to small discrepancies in planetary motion, despite its similar predictive accuracy to the geocentric model. 
This highlights the inherent uncertainties in data-driven model inference and the scientific importance of selecting physically meaningful solutions. 
For example, consider applying symbolic regression~\citep{vladislavleva2008order}, which enables the interpretation of complex machine learning models by expressing the input--output relationship of the acquired machine learning model in terms of elementary functions.
Such uncertainty runs the risk of giving a wrong physics interpretation of the data, as discussed below. 
We use historical specific examples to illustrate the importance of this issue. 
Given the observational data before Kepler, both geocentric and heliocentric models of the era were comparable in terms of predictive performance~\citep{principe2011scientific}.
In other words, when model selection was based on predictive performance, there was an uncertainty that the model was not uniquely determined. 
When machine learning was applied to such observational data, it could potentially provide a heliocentric model. 
If we extract the interpretable information from this trained machine learning model, it would lead to the conclusion that the heliocentric hypothesis was correct (Fig.~\ref{fig1}), and if this were believed, the history of science would have been different from what it is today. 
In other words, if scientific research is inadvertently conducted with machine learning, there is a risk of drawing physically incorrect conclusions. \par
In general, uncertainties in data-driven modeling can be of three types, arising from three distinct sources~\citep{Pelz2021}. 
The first type is the ``structural uncertainty'', which persists even with infinite, noiseless data when multiple model structures are consistent with the same observations. 
The second is the ``model-form uncertainty'', which emerges when the true physical law lies outside the assumed model class, leading to model-mismatch-induced indeterminacy. 
The third is the ``data uncertainty'', caused by finite or noisy data even for identifiable models, which imposes uncertainties on inferred structures. 
Note that the finiteness of data inherently introduces uncertainties, even in the absence of noise, when the degrees of freedom of the true model exceed the sample size. 
The simplest example is that a linear function cannot be determined from a single sample.
Also note that the second and third types of uncertainty encompass statistical indeterminacy, i.e., statistical identifiability~\citep{Pelz2021, rene2025selecting}. 
The approach to dealing with these uncertainties differs depending on which type exists.
In the case of the structural uncertainty, the uncertainty is an inherent structure of the physical system, and if a unique solution is desired, new physical constraints must be imposed.
Furthermore, understanding the structure of this uncertainty is important, so that, for example, the uncertainty of symmetry in Hamilton systems can be linked to conservation laws.
In the case of the model-form uncertainty, the model should be improved.
In the case of the data uncertainty, the uncertainty should be modeled through statistical modeling, etc.
Specifically, one possibility is to evaluate the posterior distribution after performing Bayesian modeling. 
Thus, distinguishing these types of uncertainty is essential for assessing the reliability of data-driven physical inference. 
Among these three types of uncertainty, structural uncertainty can be evaluated by analyzing the model itself applied to the data. 
Therefore, it should be assessed prior to evaluating other types of uncertainty.\par
One of the most commonly used forms of physical models is the partial differential equations (PDEs). 
In such PDEs, the coefficient functions are particularly crucial elements governing physical systems and often constitute the essence of physical modeling. 
In the canonical equations of motion of the Hamiltonian system, the coefficient function corresponds to the Hamiltonian function, and the observable function to the position and momentum. 
In the field of physics-informed machine learning (PIML)~\citep{karniadakis2021physics,hao2022physics}, physics-informed neural networks (PINNs)~\citep{raissi2019physics, adams2024physics, depina2022application, sahin2024solving, yang2021b} or Hamiltonian neural networks (HNN)~\citep{schmidt2009distilling, NIPS2019_9672, toth2019hamiltonian, bondesan2019learning} introduce PDEs that follow observational data as physics constraints, and then the partial differential coefficient functions are modeled by DNNs. 
This framework enables the data-driven understanding and prediction of complex physical phenomena. 
On the other hands, coefficient functions in PIML are typically estimated on the basis of prediction On the other hand, since coefficient functions in PIML are typically estimated on the basis of prediction performance, it may sometimes lead to uncertainties for the estimation of the partial differential coefficient function. 
As discussed first paragraph in this section, it is dangerous to use such an estimation model in PIML to make a physics interpretation of the phenomena. 
Since a structural uncertainty can be evaluated by analyzing the model itself, it would be useful to first assess the structural uncertainty of the coefficient functions of the PDEs in PIML. 
We note again that the ``structural uncertainty'' refers to the inherent uncertainty of the physical model itself, which specifically arises from the uncertainty in the coefficient function of the partial differential equation—not from the machine learning model or its training process.\par
To develop collaborations between scientists and machine learning to obtain a proper understanding of natural phenomena, it is necessary to evaluate what structural uncertainties exist in the data-driven estimation of physical models and partial differential coefficient functions before applying machine learning. 
The ill-posedness of inverse problems—particularly the lack of uniqueness—has been studied extensively in relation of mathematical inverse theory, such as Calderón's problem~\citep{calderon1980inverse}, whose resolution in the elliptic case~\citep{sylvester1987global} established uniqueness under full-boundary measurement assumptions. 
For hyperbolic equations, Carleman estimates have proven instrumental in deriving conditional uniqueness and stability results under geometric constraints~\citep{yamamoto2009carleman, bellassoued2017carleman}. 
In the context of machine learning approaches, Krishnapriyan et al.~\citep{krishnapriyan2021characterizing} demonstrated that PINNs may converge to physically incorrect solutions even when loss values are small, owing to flat or multimodal optimization landscapes. Mishra et al.~\citep{mishra2022estimates} further analyzed such failure modes via uncertainty quantification, highlighting the epistemic structural uncertainty inherent in inverse modeling without proper constraints. 
These studies focus only on specific classes of PDEs and, therefore, cannot be used to evaluate the structural uncertainty of the more diverse classes of PDEs used in physical research. 
If a framework for evaluating the structural uncertainty in estimating coefficient functions can be constructed for a wide class of PDEs, it will facilitate the promotion of effective physics research using machine learning. \par
The purpose of this study is to develop a framework to quantitatively evaluate the degree of structural uncertainty and its structure in the inductive estimation of coefficient functions for a wide range classes of PDEs.

\section{Related Works}
\label{sec_2}
\subsection{PIML and Inverse Problem}
PIML is an emerging framework that integrates physical laws, such as PDEs, into the training process of machine learning models such as DNNs. 
Prominent examples of instantiation of this idea are the Physics-Informed Neural Networks (PINNs)~\citep{raissi2019physics} and Hamiltonian neural networks (HNN)~\citep{greydanus2019hamiltonian}, where a neural network is trained not only to fit observational data but also to satisfy a given PDE.\par
In the PINNs framework, the governing PDE is typically of the form
\[
\mathcal{N}[u ,a](x) = 0, \quad x\in \mathbb{R}^d \times [0,T],
\]
where \( \mathcal{N} \) is a nonlinear differential operator derived from physical laws, \( u : \mathbb{R}^d \to \mathbb{R} \) is a sufficiently smooth function of observation values, \( a : \mathbb{R}^d \to \mathbb{R} \) is a sufficiently smooth function of coefficients of PDE, and $x := (x,t) \in \mathbb{R}^d \times [0,T]$. 
For example, the following equations are included: $\mathcal{N}[u, a] := u(x) - \nabla \cdot (a(x) \nabla u(x)) = 0$.
To enforce this PDE constraint in machine learning, the loss function used in training is augmented by a physics-informed term:
\begin{eqnarray}
\label{eq_pinns}
\mathcal{L}(\theta_u,\theta_a) = \mathcal{L}_{\text{data}}(\theta_u) + \lambda_{\text{PDE}} \mathcal{L}_{\text{PDE}}(\theta_u,\theta_a),
\end{eqnarray}
where
\begin{eqnarray}
\mathcal{L}_{\text{data}}(\theta_u) = \frac{1}{N} \sum_{i=1}^{N} \left\lVert u_\theta(x_i) - u_i \right\rVert^2,
\quad
\mathcal{L}_{\text{PDE}}(\theta_u,\theta_a) = \frac{1}{N_r} \sum_{r=1}^{N_r} \left\lVert \mathcal{N}[u_{\theta_u}, a_{\theta_a}](x_r) \right\rVert^2,
\end{eqnarray}
where \( u_{\theta_u}(x) \) and \( a_{\theta_a}(x) \) are a neural network models parametrized by $\theta$. 
Here, \( \{(x_i, u_i)\}_{i=1}^{N} \) are supervised data points, wherease \( \{(x_r)\}_{r=1}^{N_r} \) are residual points where the PDE is enforced. 
The parameter \( \lambda_{\text{PDE}} \) balances the relative importance of data fidelity and physics conformity. 
In this framework, the partial differential coefficient function $a(x,t)$ can also be estimated by minimizing $\mathcal{L}(\theta_u,\theta_a)$.  
In the HNN-type framework, the objective is not to estimate the observation function $u(x)$, but to estimate the partial differential coefficient function $a(x)$. 
Thus, the loss function is given as follows. 
\begin{eqnarray}
\mathcal{L}_{\text{HNN}}(\theta) = \frac{1}{N} \sum_{i=1}^{N} \left\lVert \mathcal{N}[u, a_{\theta}](x_i) \right\rVert^2,
\end{eqnarray}
where \( \{(x_i, u_i)\}_{i=1}^{N} \) are given as supervised data. From there, the partial derivatives of $u(x)$ in PDE, $\mathcal{N}[u, a_{\theta}](x_i)$, are assumed to be given numerically. 
For example, if the PDE is a canonical equation of motion, the loss function is given by
\begin{equation}
\label{eq:hnn-loss}
\mathcal{L}_{\text{HNN}}(\theta) = \frac{1}{N} \sum_{i=1}^{N} \left\|
\begin{bmatrix}
\frac{\partial H_\theta}{\partial p}(q_i, p_i) \\
- \frac{\partial H_\theta}{\partial q}(q_i, p_i)
\end{bmatrix}
-
\begin{bmatrix}
\dot{q}_i^{\text{obs}} \\
\dot{p}_i^{\text{obs}}
\end{bmatrix}
\right\|^2,
\end{equation}
where the observation function is $u(t,q,p) = (t,q,p)$ and the coefficient function is $a(t,q,p) = H(q,p) $. 
In HNNs, the coefficient function $a(x)$ is estimated by minimizing the loss function $\mathcal{L}_{\text{HNN}}(\theta)$ similar to that of PINNs.\par
PIML has been successfully demonstrated in various tasks, including forward simulation, spatiotemporal forecasting \citep{karniadakis2021physics}, and inverse problems such as parameter estimation \citep{raissi2018hidden}. 
For inverse problems, the physical constraint often compensates for limited data, enabling the estimation of unknown coefficient functions or source terms. 
However, the learned solution may not be unique: the PDE residual can be small even when multiple, distinct functions explain the data equally well using the same physical model.\par
Recent studies have highlighted the lack of identifiability guarantees in PINNs \citep{yang2021b}. 
In particular, the minimization of \( \mathcal{L}_{\text{PDE}} \) does not necessarily imply that the estimated parameters or functions are physically meaningful or unique. 
Furthermore, the structure of the differential operator \( \mathcal{N} \), the available observation data, and the expressivity of the neural network all affect the identifiability and uncertainty of the learned solution. These findings emphasize the need for a rigorous theoretical framework for understanding the ill-posedness and uncertainty inherent in physics-informed inverse problems.

\subsection{Uncertainty in Inverse Problems}

There are only a few studies that mathematically analyze the degree of uncertainty and its structure in the inductive estimation of coefficient functions of PDEs, and these studies are limited to specific classes of PDEs. 
These studies are introduced as follows. 
Classical studies such as Calderón's problem~\citep{calderon1980inverse} and its resolution in the elliptic case~\citep{sylvester1987global} established the uncertainty evaluation of the coefficient function under full-boundary measurement assumptions. 
For hyperbolic equations, Carleman estimates have proven instrumental in deriving conditional uniqueness and stability results under geometric constraints~\citep{yamamoto2009carleman, bellassoued2017carleman}. 
Thus, in the theoretical approach, analysis is limited to a specific class of PDEs.\par
In the context of machine learning approaches, although there have been studies examining the presence or absence of uncertainties in coefficient functions or the qualitative degree of an uncertainty, there is no framework for quantitatively evaluating the specific degree of the uncertainty or its structure. These studies are introduced as follows. 
Krishnapriyan et al.~\citep{krishnapriyan2021characterizing} demonstrated that PINNs may converge to physically incorrect solutions even when loss values are small, owing to flat or multimodal optimization landscapes. 
Mishra et al.~\citep{mishra2022estimates} further analyzed such failure modes by uncertainty quantification, highlighting the epistemic uncertainty inherent in inverse modeling without proper constraints.\par
Bayesian extensions, such as B-PINNs~\citep{yang2021b, mishra2022estimates}, provide a qualitative uncertainty evaluation by placing distributions over unknowns and inferring posteriors via variational or sampling-based methods. 
Although it might be possible to use information of posterior distributions (e.g., their unimodality or variance) from Bayesian PINNs to indirectly evaluate uncertainties and consider candidate constraints, this would still require threshold criteria (e.g., a threshold of posterior variance to decide it as an identifiable distribution) to decide whether a parameter is determined or not. 
Such thresholds are not defined in the existing Bayesian PINN literature.
Thus, there has been no research in which the degree of uncertainty and its structure have been quantitatively evaluated in the inductive estimation of coefficient functions for a wide range of classes of PDEs.

\section{Theoretical Preliminaries for Uncertainty Evaluation}
\label{sec_3}

\begin{dfn}[$k$-Leaf Set of Partial Derivatives]
Let \( a : \Omega \to \mathbb{R} \) be a sufficiently smooth function defined on an open domain \( \Omega \subset \mathbb{R}^d \).  
Using the multi-index notation \( \alpha = (\alpha_1, \dots, \alpha_d) \in \mathbb{N}^d \), we obtain the arbitrary partial differential coefficient of $a$ as
\[
|\alpha| := \sum_{i=1}^d \alpha_i, \quad
\partial^\alpha a(x) := \frac{\partial^{|\alpha|} a}{\partial x_1^{\alpha_1} \cdots \partial x_d^{\alpha_d}}.
\]
Let \( A_k := \{ \alpha \in \mathbb{N}^d \mid |\alpha| = k \} \) be the set of all multi-indices of the total order \(k\).  
Then, the \emph{$k$-leaf set of partial derivatives} $S^{\rm leaf}_k$ is defined as 
\[
S^{\rm leaf}_k := \left\{ \partial^\alpha a(x)\partial^\beta a(x) \mid \alpha \in A_k, |\beta| =  1\right\}.
\]\par
Example: If $\{\partial^\alpha a(x) \mid \alpha \in A_1\}$ is given by $\{\partial_{x}a(x),\partial_{y}a(x)\}$, then the $2$-leaf set of partial derivatives is $\left\{ \partial_{xx}a(x),\partial_{xy}a(x),\partial_{yx}a(x),\partial_{yy}a(x) \right\}$. 
\end{dfn}
 \par

\begin{thm}[Uniqueness of Coefficient Function up to Polynomial under Root Derivative Information]
\label{theorem1}
Let \( u : \mathbb{R}^d \to \mathbb{R} \) and \( a : \mathbb{R}^d \to \mathbb{R} \) be a sufficiently smooth function, and consideration a $m$-th PDEs of the form
\[
\sum_{\alpha \in A_{\geq m}} \varphi^{(\ell)}_\alpha(x, u(x), \partial u(x), \partial^2 u(x), \dots) \cdot \partial^\alpha a(x) = C^{(\ell)}, C:{\rm const.},\quad \ell = 1, \dots, L,
\]
where $A_{\geq m} := \{\alpha \mid m \leq |\alpha|\}$, $m \geq k$, and $ \partial^k u(x)$ represents the arbitrary set of $k$-th-order partial differential coefficients. 
That is, PDEs is linear in \( \partial^\alpha a(x) \). \\
Assume that PDEs have a $k$-leaf set $S^{\rm leaf}_k$ in their equations. 
Discretize \(\Omega\) on an infinitesimal grid with spacing \(\varepsilon > 0\), and denote the grid points as \( x^{(i)} \in \mathbb{R}^d \) for \( i = 1, \dots, N \).  
For each grid point, consider the discretized PDE system:
\[
\sum_{\alpha \in A_{\geq m}} \varphi^{(\ell)}_{\alpha} \left( x_{(i)}, u(x_{(i)}), u^2(x_{(i)}), \partial u(x_{(i)}), \dots \right) \cdot \partial^{\alpha} a(x_{(i)}) = C_{(i)}^{(\ell)}, \quad \ell = 1, \dots, L.
\]
Then, by stacking the equations across all grid points, the system is represented as a linear system:
\[
\mathbf{M} \cdot \mathbf{a} = \mathbf{c},
\]
where \(\mathbf{a} \in \mathbb{R}^{|A_{\leq m}| \cdot N}\) is the vector of derivatives of the coefficient  function $a(x)$, \(\mathbf{M} \in \mathbb{R}^{LN \times |A_{\leq m}|N}\) is the matrix constructed from \(\varphi^{(\ell)}_{\alpha}\), and \(\mathbf{c} \in \mathbb{R}^{|A_{\leq m}| \cdot N}\) is the vector of constant values. 
Then, the following can be stated:
\begin{itemize}
\item If \({\rm rank}(\mathbf{M}) = {\rm rank}(\mathbf{M},\mathbf{c}) = |A_{\leq m}|N\) , then the coefficient function \(a(x)\) is uniquely determined up to a polynomial of degree of at most \(k-1\): \[
\tilde{a}(x) = a(x) + p(x),
\]
where \( p(x) \) is a polynomial of total degree of at most \( k - 1 \).
\end{itemize}

\end{thm}

{\bf Proof}\\
If \({\rm rank}(\mathbf{M}) = {\rm rank}(\mathbf{M},\mathbf{c}) = |A_{\leq m}|N\), the $k$-leaf set of derivatives $S^{\rm leaf}_k$ is determined on arbitrarily position $x$. 
In other words, the $k$-th-order partial differential coefficients of $a(x)$ are uniquely determined on an infinitesimal small-spaced grid. 
Then, it is shown that the partial differential coefficient function $a(x)$ is uniquely determined except for the uncertainty of the $k-1$ degree polynomial as follows.
Consider the case $k=0,\:d=2$. 
If $x=(x,y)$, then
\[
a(x,y) - a(x_0,y_0) = \left[a(x,y_0) - a(x_0,y_0)\right] + \left[a(x,y) - a(x,y_0)\right].
\]
Applying the fundamental theorem of calculus to the right-hand side, we obtain
\[
a(x,y) - a(x_0,y_0) = \int_{x_0}^x dx \partial_x a(x,y_0) + \int_{y_0}^{y} dy \partial_y a(x,y).
\]
Furthermore, from the definition of integral, it can be transformed as follows.
\[
a(x,y) - a(x_0,k_0)= \lim_{\Delta_x\rightarrow 0}\sum_{i=1}^{n_x} \partial_x a(x_0+i\Delta_x,y_0) \Delta_x  + \lim_{\Delta_y\rightarrow 0}\sum_{j=1}^{n_k} \partial_y a(x,y_0+j\Delta_k) \Delta_y
\]
Because the $k$-th-order partial differential coefficients of $a(x)$ are uniquely determined on an infinitesimal small--spaced grid, the right-hand side can be calculated.
Thus, it is shown that the partial differential coefficient function $a(x,y)$ at arbitrary coordinates $(x,y)$ can be uniquely estimated except for the uncertainty of the constant $a(x_0,k_0)$. 
For general $k$ and $d$, we also decompose the expression as:
\begin{eqnarray}
\partial^k a(x_1, x_2, \ldots, x_d)
- \partial^k a(x_1^{(0)}, x_2^{(0)}, \ldots, x_d^{(0)})
&=&
\Big[
\partial^{k+1} a(x_1, x_2^{(0)}, \ldots, x_d^{(0)})
- \partial^{k+1} a(x_1^{(0)}, x_2^{(0)}, \ldots, x_d^{(0)})
\Big]
\nonumber\\
&& {} + \cdots +
\Big[
\partial^{k+1} a(x_1, \ldots, x_{d-1}, x_d)
- \partial^{k+1} a(x_1, \ldots, x_{d-1}, x_d^{(0)})
\Big].\nonumber
\end{eqnarray}

Then, each term transforms to an integral form in a same manner as the case of $k=0, d=2$, completing the inductive argument.
\hfill $\blacksquare$
\par

Note that if \({\rm rank}(\mathbf{M}) \neq  |A_{\leq m}|N\), but the $k$-leaf set of derivatives is uniquely determined, then the coefficient function \(a(x)\) is also uniquely determined up to a polynomial of degree of at most \(k-1\). 
In this case, the proofs and proposed methods can be set up in the same manner. 
Also, even if some $k$th-order partial differential coefficients are undefined or not included in the PDEs, the same argument holds if the corresponding lower-order partial differential coefficients can be estimated.

\subsection{Examples}
\label{sec_examples}
 {\bf $\bullet$ Hamiltonian system}\\
    Let the canonical variables be denoted by
$(\boldsymbol{q}, \boldsymbol{p})^\top \in \mathbb{R}^{2d}$,
where $\boldsymbol{q} = (q_1, \ldots, q_d)^\top$ are the generalized coordinates and $\boldsymbol{p} = (p_1, \ldots, p_d)^\top$ are the generalized momenta.

Given the Hamiltonian function $H(\boldsymbol{q}, \boldsymbol{p})$, the canonical equations of motion (Hamilton's equations) can be expressed in matrix form as
$\begin{pmatrix}
\frac{d \boldsymbol{q}}{dt} \\
\frac{d \boldsymbol{p}}{dt}
\end{pmatrix} =: 
\begin{pmatrix}
\dot{\boldsymbol{q}} \\
\dot{\boldsymbol{p}}
\end{pmatrix} =
\begin{pmatrix}
\boldsymbol{0} & I_n \\
- I_n & \boldsymbol{0}
\end{pmatrix} \begin{pmatrix}
\frac{\partial H}{\partial \boldsymbol{q}} \\
\frac{\partial H}{\partial \boldsymbol{p}}
\end{pmatrix}$. 
The equation on an infinitesimal $N$ grid space is written as $\mathbf{M} \cdot \mathbf{a} = \mathbf{c}$, where $M = \begin{pmatrix}
\boldsymbol{0} & I_{Nd} \\
- I_{Nd} & \boldsymbol{0}
\end{pmatrix}$, 
$\mathbf{a} = \begin{pmatrix}
\frac{\partial H}{\partial \boldsymbol{q}_1},
\dots,
\frac{\partial H}{\partial \boldsymbol{q}_N},
\frac{\partial H}{\partial \boldsymbol{p}_1},
\dots,
\frac{\partial H}{\partial \boldsymbol{p}_N}
\end{pmatrix}^\top$, and $\mathbf{c}=
\begin{pmatrix}
\dot{\boldsymbol{q}}_1,
\dots,
\dot{\boldsymbol{q}}_N,
\dot{\boldsymbol{p}}_1,
\dots,
\dot{\boldsymbol{p}}_N
\end{pmatrix}^\top$. 
Since ${\rm rank}(\boldsymbol{M}) = 2Nd$, the necessary conditions are satisfied such that the Hamiltonian function $H(\boldsymbol{q},\boldsymbol{p})$ is uniquely determined, except for the indefiniteness of the constant. \\
 \\
 \\

 {\bf $\bullet$ Lagrange system}\\
Let the generalized coordinates be denoted by $\boldsymbol{q} = (q_1,\dots, q_d)^\top$.
Lagrange's equations of motion can be written in matrix form as 
$
\dot{\boldsymbol{p}} := \frac{d}{dt} \frac{\partial L}{\partial \dot{\boldsymbol{q}}} =
\begin{pmatrix}
\boldsymbol{0} & I_d
\end{pmatrix} \begin{pmatrix}
\frac{\partial L}{\partial \boldsymbol{q}} \\
\frac{\partial L}{\partial \dot{\boldsymbol{q}}}
\end{pmatrix}.
$
The equation on an infinitesimal $N$ grid space is written as $\mathbf{M} \cdot \mathbf{a} = \mathbf{c}$, where 
$\mathbf{M}=
\begin{pmatrix}
\boldsymbol{0} & I_{Nd}
\end{pmatrix}$, 
$\mathbf{a} =
\begin{pmatrix}
\frac{\partial L}{\partial \boldsymbol{q}_1},
\dots,
\frac{\partial L}{\partial \boldsymbol{q}_N},
\frac{\partial L}{\partial \dot{\boldsymbol{q}}_1},
\dots,
\frac{\partial L}{\partial \dot{\boldsymbol{q}}_N},
\end{pmatrix}^{\top}$, and 
$\mathbf{c} =
\begin{pmatrix}
\dot{\boldsymbol{p}}_1,
\dots,
\dot{\boldsymbol{p}}_N
\end{pmatrix}^{\top}$. 
Since ${\rm rank}(\boldsymbol{M}) = Nd < 2Nd$, the Lagrange function $L(\boldsymbol{q},\dot{\boldsymbol{q}})$ is undetermined.\par
Since the Hamiltonian and Lagrangian systems have a transformable relationship through the Légendre transformation, it seems counterintuitive that only the Lagrangian is not indefinite.
The reason the Lagrangian cannot be determined is that the information corresponding to the part of the canonical equation of motion in the Hamiltonian system, $\dot{\boldsymbol{q}} := \frac{\partial H}{\partial \boldsymbol{p}}$, is missing in the Lagrangian system. 
Since one physical constraint for estimating the coefficient function has disappeared, the Lagrange function is not determined.
This missing information corresponds to the definition of the generalized momentum in the Lagrangian system, $\boldsymbol{p} := \frac{\partial L}{\partial \dot{\boldsymbol{q}}}$. 
In fact, adding the definition of generalized momentum to the Lagrangian equation of motion leads to the satisfaction of the necessary condition, ${\rm rank}(\boldsymbol{M}) = 2Nd$, for the Lagrangian to be uniquely determined.

\section{Proposed Framework: PIML with Uncertainty Evaluation}
As discussed in Sec.~\ref{sec_2}, when estimating the coefficient function $a(x)$ using PIML, uncertainty in the physics system results in physically inappropriate learning. 
Under the mathematical preparation in Sec.~\ref{sec_3}, we propose a three-step framework for obtaining a scientifically valid model in the PIML framework. 
\begin{enumerate}
\item[{\bf step 1}] Before considering the implementation of the machine learning model, do the following. By evaluating the rank of $\boldsymbol{M}$ (in Theorem~\ref{theorem1}), we acquire information on the degree of uncertainty and its structure of the coefficient function in a given partial differential equation. 
\item[{\bf step 2}] Depending on the structure of $\boldsymbol{M}$, introduce physical constraints to the loss function (Eq.~(\ref{eq_pinns}) or (\ref{eq:hnn-loss})) that reduce the uncertainty. 
The machine learning model will be trained using the loss function.
\item[{\bf step 3}] Examine how the estimation results of $a(x)$ change with the change in strength of the physical constraints.
\end{enumerate}
{\bf Step 3} states that the proposed framework does not estimate the hyperparameters of machine learning, unlike in the case of common machine learning.
This is because it is risky to determine the strength of a given physical constraint on the basis of solely its predictive performance, especially when the knowledge of what is being analyzed is unclear.
Providing the physicist with all the estimation results under all strength constraints will enable better physics interpretation.
For example, Kepler found the law of elliptical orbits by focusing on the slight deviation of Mars' orbit from a circular orbit.

\section{Demonstration}
\begin{figure}
\centering
\includegraphics[width=\linewidth]{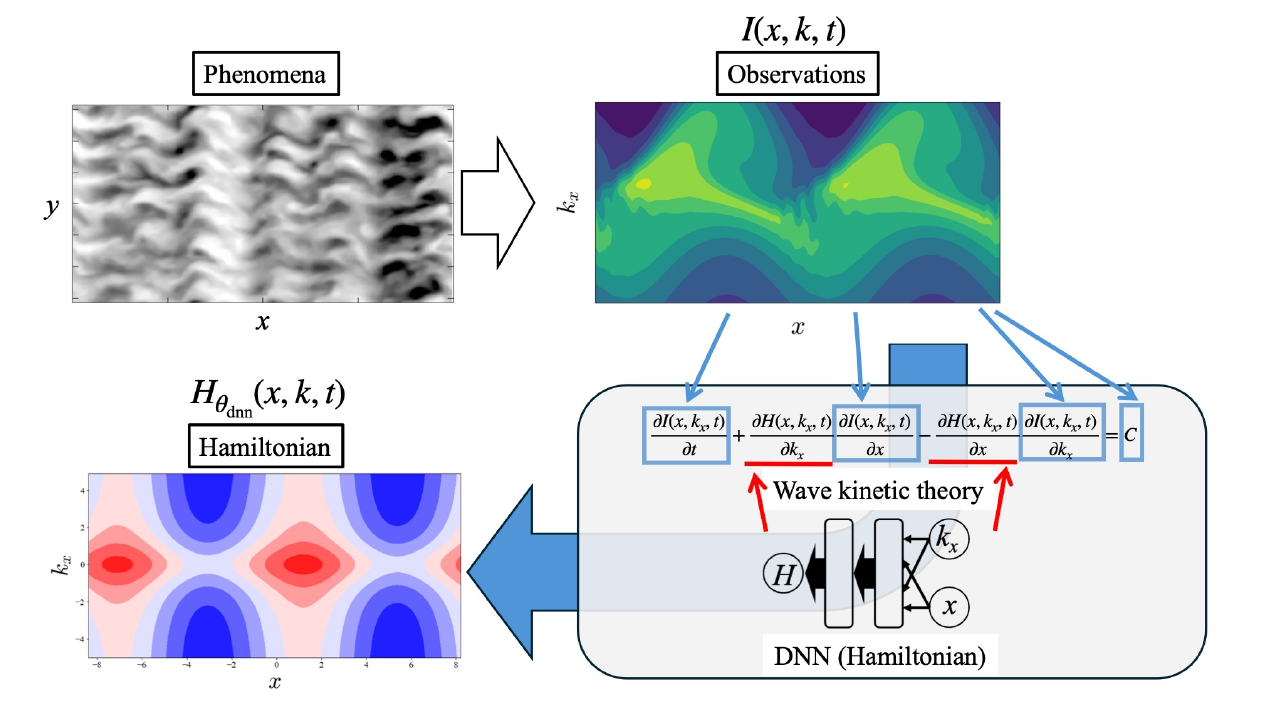}
\caption{Conceptual diagram of Hamiltonian function estimation based on the wave kinetic equation.}
\label{fig2}
\end{figure}

The proposed framework was applied to the problem of data-driven Hamiltonian function estimation for the wave kinetic equation (see Fig.~\ref{fig2}), which is important for nuclear fusion research, and the effectiveness of our framework was verified.

\subsection{Wave Kinetic Equation}
Modeling the dynamics of turbulent vortices, which emerge in complex, high-dimensional turbulence phenomena observed in fusion reactors, using low-dimensional Hamiltonian dynamical systems, such as wave kinetic equation, is useful for the prediction and control of turbulence based on physics understanding~\citep{diamond2005zonal, gurcan2015zonal, kaw2001coherent, sasaki2017enhancement, sasaki2018spatio, garbet2021zonal, sasaki2021formation}. The wave kinetic equation describes the time evolution of the density distribution function \( I(x, k_x, t) \) in the turbulence phase space \((x, k_x)\) and is given by
\begin{eqnarray}
\frac{\partial I(x,k_x,t)}{\partial t}
+ \frac{\partial H(x,k_x,t)}{\partial k_x} \frac{\partial I(x,k_x,t)}{\partial x}
- \frac{\partial H(x,k_x,t)}{\partial x} \frac{\partial I(x,k_x,t)}{\partial k_x}
= C(x,k_x,t). \label{Ex_BoltzmanEq}
\end{eqnarray}
This equation is mathematically analogous to the Boltzmann equation. Here, the term \( C(x, k_x, t) \) represents the generation and damping of turbulent vortices, and is modeled using the linear growth rate \( \gamma_L \) and the nonlinear damping rate \( \Delta \omega \) as follows:
$C(x,k_x,t) := \gamma_{L}(k_x) I(x,k_x,t) - \Delta \omega [I(x,k_x,t)]^2, \:\:
\gamma_{L}(k_x) = \frac{k_y(k_x^2 + k_y^2)}{D(1 + k_x^2 + k_y^2)^3}
\exp\left(-\left(\frac{k_x}{\Delta k}\right)^2\right),$ where \( \Delta k \) characterizes the spectral width of \( I(x, k_x, t) \) in the linear regime. The Hamiltonian function \( H(x, k_x, t) \), corresponding to the distribution of turbulence intensity, is defined as
\begin{eqnarray}
H(x,k_x,t) = H_0 + \frac{k_y}{1 + k_x^2 + k_y^2} + k_y V_y(x,t). \label{H_eq}
\end{eqnarray}
The second term on the right-hand side of Eq.~(\ref{H_eq}) corresponds to the dispersion relation of drift waves, whereas the third term represents the Doppler shift induced by the zonal flow. In other words, turbulence is deformed (i.e., its spectral distribution is changed) owing to spatially non-uniform Doppler shifts induced by the zonal flow via the third term on the left-hand side of Eq.~(\ref{Ex_BoltzmanEq}).\par
Next, we focus on the Geodesic Acoustic Mode (GAM), an oscillatory branch of zonal flows in toroidal plasmas \citep{dawson1968geodesic}. The evolution equation for GAM is given by \citep{sasaki2018spatio} as follows:
\begin{eqnarray}
\frac{\partial^2 V_y(x,t)}{\partial t^2} + \omega_G^2 V_y(x,t)
= \frac{\partial}{\partial t} \frac{\partial^2}{\partial x^2}
\int dk_x \frac{k_x k_y I(x,k_x,t)}{(1 + k_x^2 + k_y^2)^2} 
+ \mu \frac{\partial}{\partial t} \frac{\partial^2 V_y(x,t)}{\partial x^2}, \label{GAM_eq}
\end{eqnarray}
where \( \omega_G \) is the GAM frequency. The first term on the right-hand side represents the GAM driving term due to Reynolds stress, which is a functional of the turbulent phase--space distribution \( I(x, k_x, t) \). The turbulence and zonal flows are thus coupled via the third term in Eq.~(\ref{Ex_BoltzmanEq}) and the first term on the right-hand side of Eq.~(\ref{GAM_eq}).\par
In the following analysis, we use numerical solutions of the coupled Eqs.~(\ref{Ex_BoltzmanEq}) and (\ref{GAM_eq}). 
The parameters used in the simulations are 
$ k_y = 1, \; D = 3, \; \Delta k = 3, \; \omega_G = 0.1061, \;$ and $\mu = 0.05 $. 
This simulation provides the value of $I(x,k_x,t)$ on a grid in the $(x,k_x)$ space.
By computing the numerical derivatives from this simulation data, we can obtain the following dataset $DS$ with the sample size $M_xM_{k_x}$:  
$DS := \left\{\left. \partial_t I(x_i,k_j,t),  \partial_x I(x_i,k_j,t),\partial_k I(x_i,k_j,t), C(x_i,k_j,t)\right| i\in [0,\:M_x],j\in [0,\:M_{k_x}],t=\tau\right\}$.  
Note that in this demonstration, for simplicity, the time slice of the Hamiltonian, $H(x,k_x,t=\tau)$, is estimated independently at each time $\tau$.\par

\begin{wrapfigure}{r}[0pt]{0.6\textwidth}
\centering
\includegraphics[width=0.6\textwidth]{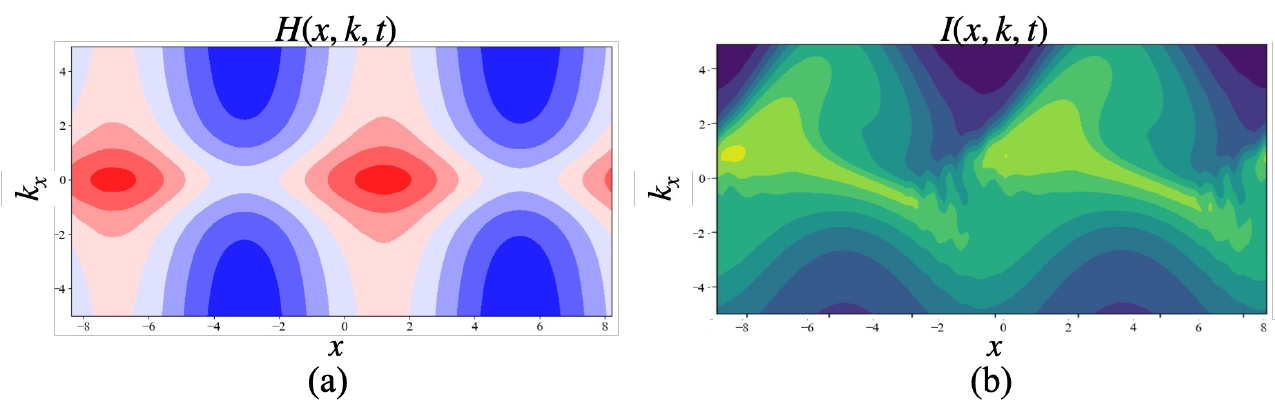}
\caption{(a) Hamiltonian function $H(x,k_x)$ set up in the simulation. (b) Turbulence intensity data $I(x,k_x)$ obtained from the simulation.}
\label{fig3}
\end{wrapfigure}

The objective of this analysis is to inductively estimate the Hamiltonian function \( H(x, k_x, t) \) from the observational data \( I(x, k_x, t) \). 
In understanding the mechanisms of turbulent phenomena based on the coarse-grained wave kinetic equation, a key bottleneck lies in establishing the correspondence of its simulation results to real-world phenomena.  
Traditionally, this correspondence is achieved through the manual design of Hamiltonians by scientists based on their insights into physical phenomena. However, designing a Hamiltonian that accurately reflects complex real-world phenomena—affected by various factors—is generally a challenging task.
To assist scientists in the design of such Hamiltonians, we aim to develop a data-driven framework for Hamiltonian estimation.
Specifically, we attempt to inversely estimate the Hamiltonian function from measurement data using a Hamiltonian neural network (HNN)-based approach.
If successful, this inverse estimation would enable the extraction of physically meaningful information from DNN and provide valuable support for scientists engaged in Hamiltonian modeling.

\subsection{Uncertainty Evaluation ({\bf Step 1})}
Given the turbulence intensity function $I(x,k_x,t)$, we perform an uncertainty evaluation when estimating the Hamiltonian $H(x,k_x,t)$ that the turbulence follows under the constraints of the wave kinetic equation [Eq.~(\ref{Ex_BoltzmanEq})].
First, the wave kinetic equation is expressed in the infinitesimally small-interval $N_x\times N_{k_x}=\infty \times \infty$ grid space as follows.
\begin{eqnarray}
\left(\begin{matrix}
I_t(x^{(1)},k_x^{(1)}) - C(x^{(1)},k_x^{(1)},\tau)\\
I_t(x^{(1)},k_x^{(2)}) - C(x^{(1)},k_x^{(2)},\tau)\\
\vdots\\
I_t(x^{(N_x)},k_x^{(N_{k_x}-1)}) - C(x^{(N_x)},k_x^{(N_{k_x}-1)},\tau)\\
I_t(x^{(N_x)},k_x^{(N_{k_x})}) - C(x^{(N_x)},k_x^{(N_{k_x})},\tau)
\end{matrix}\right)=
\mathbf{M}
\left(\begin{matrix}
H_x(x^{(1)},k_x^{(1)})\\
H_{k_x}(x^{(1)},k_x^{(1)})\\
\vdots\\
H_x(x^{(N_x)},k_x^{(N_{k_x})})\\
H_{k_x}(x^{(N_x)},k_x^{(N_{k_x})})
\end{matrix}\right),
\end{eqnarray}
\begin{eqnarray}
\mathbf{M} := \left(\begin{matrix}
-I_{k_x}(x^{(1)},k_x^{(1)})&I_x(x^{(1)},k_x^{(1)})&0 &0 & & & \\
0& 0& -I_{k_x}(x^{(1)},k_x^{(2)})&I_x(x^{(1)},k_x^{(2)})& & & \\
 & & & & \ddots\\
 & & & & \\
\end{matrix}\right),\nonumber
\end{eqnarray}
\begin{eqnarray}
H_z(x^{(i)},k_x^{(j)}):=\left.\frac{\partial H(x,k_x,t)}{\partial z}\right|_{x=x^{(i)},k=k_x^{(j)},t=\tau},
I_z(x^{(i)},k_x^{(j)}):=\left.\frac{\partial I(x,k_x,t)}{\partial z}\right|_{x=x^{(i)},k=k_x^{(j)},t=\tau},\nonumber
\end{eqnarray}
where $z\in\{x,k_x,t\}$, and matrix size of $\mathbf{M}$ is $N_xN_{k_x} \times 2N_xN_{k_x}$.
We can see that ${\rm rank}(\mathbf{M}) = N_xN_{k_x} < 2N_xN_{k_x}$ and that is why the solution is undefined, and that $N_xN_{k_x}$ of PDEs are not enough to determine the Hamiltonian function uniquely. \par

\subsection{Introduce Physical Constraints ({\bf Step 2})}
The uncertainty of the Hamiltonian estimation is avoided by introducing physical constraints.
Assuming now that there is no anisotropy in the $x$ direction in the motion of the system, the Hamiltonian function is line symmetric centered at $k_x=0$.
In fact, the Hamiltonian function used in data generation has line symmetry centered at $k_x=0$ [Fig.~\ref{fig3}(a)].
This constraint implies that $H\left(x^{(i)},-k_x^{(j)},t^{(k)}\right) = H\left(x^{(i)},k_x^{(j)},t^{(k)}\right)$, $\partial_x H\left(x^{(i)},-k_x^{(j)},t^{(k)}\right) = \partial_x H\left(x^{(i)},k_x^{(j)},t^{(k)}\right)$, and $\partial_k H\left(x^{(i)},-k_x^{(j)},t^{(k)}\right) = -\partial_k H\left(x^{(i)},k_x^{(j)},t^{(k)}\right)$. 
Introducing this constraint into the wave kinetic equation on the grids gives the following representation with block matrices.
\begin{eqnarray}
\left(\begin{matrix}
I_t(x^{(1)},k_x^{(1)}) - C(x^{(1)},k_x^{(1)})\\
I_t(x^{(1)},-k_x^{(1)}) - C(x^{(1)},-k_x^{(1)})\\
\vdots\\
I_t(x^{(N_x)},k_x^{(N_{k_x}/2)}) - C(x^{(N_x)},k_x^{(N_{k_x}/2)})\\
I_t(x^{(N_x)},-k_x^{(N_{k_x}/2)}) - C(x^{(N_x)},-k_x^{(N_{k_x}/2)})
\end{matrix}\right)
=\boldsymbol{M}
\left(\begin{matrix}
H_x(x^{(1)},k_x^{(1)})\\
H_{k_x}(x^{(1)},k_x^{(1)})\\
\vdots\\
H_x(x^{(N_x/2)},k_x^{(N_{k_x}/2)})\\
H_{k_x}(x^{(N_x/2)},k_x^{(N_{k_x}/2)})
\end{matrix}\right),
\end{eqnarray}
\footnotesize
\begin{eqnarray}
\boldsymbol{M}=
\left(\begin{matrix}
-I_{k_x}(x^{(1)},k_x^{(1)})&I_x(x^{(1)},k_x^{(1)})& & & & & \\
-I_{k_x}(x^{(1)},-k_x^{(2)})&I_x(x^{(1)},-k_x^{(2)})& & &\text{\huge{0}} & &\\
 & & &\ddots & \\
  & \text{\huge{0}}& & &-I_{k_x}(x^{(N_x/2)},k_x^{(N_{k_x}/2)})&I_x(x^{(N_x/2)},k_x^{(N_{k_x}/2)})&\\
  & & & &-I_{k_x}(x^{(N_x/2)},-k_x^{(N_{k_x}/2)})&I_x(x^{(N_x/2)},-k_x^{(N_{k_x}/2)})&\\
\end{matrix}\right),\nonumber
\end{eqnarray}
\normalsize
where the matrix size of $\mathbf{M}$ is $N_xN_{k_x} \times N_xN_{k_x}$. 
Since the number of partial differential coefficients of the unknown Hamiltonian is $N_xN_{k_x}$, if ${\rm rank}(\mathbf{M}) = N_xN_{k_x}$, the Hamiltonian function is uniquely determined, except for the uncertainty of the constant. 
For this condition to be satisfied, it must be $ \forall\:i,j,\: {\rm rank}\left[\left(\begin{matrix}
-I_k(x_i,k_j)&I_x(x_i,k_j)\\
-I_k(x_i,-k_j)&-I_x(x_i,-k_j)
\end{matrix}\right)\right] = 2$. 
This is true if the turbulence intensity distribution $I_x(x,k)$ has a gradient at all points and has no line symmetry centered at $k_x=0$.
Since this is true for the present dataset [Fig.~\ref{fig3}(b)], the Hamiltonian function is physically uniquely determined by adding the symmetry constraint, except for the uncertainty of the constant.
According to the results of the above evaluation of uncertainty, we designed the loss function as follows.
\footnotesize
\begin{eqnarray}
\label{eq_loss_wq}
{\rm Loss}(\mathbf{\theta}_{\rm dnn}) &=& \frac{1}{M_x M_{k_x}}\sum_{i,j}\left\lVert\partial_t I(x_i,k_j) - C(x_i,k_j) -\partial_x H_{\theta_{\rm dnn}}(x_i,k_j)\partial_x I(x_i,k_j)+\partial_x H_{\theta_{\rm dnn}}(x_i,k_j)\partial_k I(x_i,k_j)\right\rVert^2\nonumber\\
&+&\lambda \frac{1}{M_x M_{k_x}}\sum_{i,j}  \left\lVert H_{\theta_{\rm dnn}}(x_i,k_j) - H_{\theta_{\rm dnn}}(x_i,-k_j)\right\rVert^2.
\end{eqnarray}
\normalsize

For further details on the neural network model and other aspects, please refer to Appendix~\ref{sec_dnn_params} and the code available at the following URL: \url{https://anonymous.4open.science/r/Structural_uncertainty-30D5}.

\subsection{Estimation Results of Hamiltonian for Each Hyperparameter $\lambda$ ({\bf Step 3})}
\begin{figure}[tb]
\centering
\includegraphics[width=1.0\linewidth]{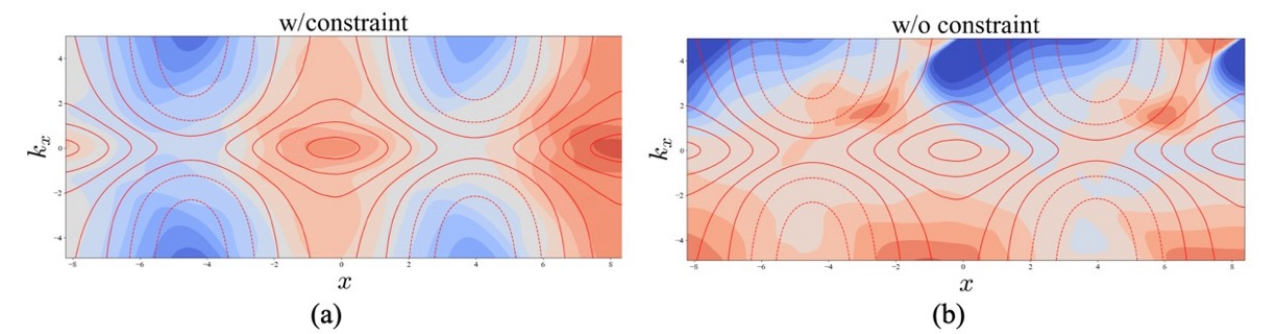}
\caption{(a) Estimation results for the Hamiltonian function $H_{\theta_{\rm dnn}}(x,k_x)$ with symmetry constraints and (b) without constraints. The histogram represents the DNN function estimation results, and the red contour line represents the Hamiltonian function set when generating the dataset.}
\label{fig4}
\end{figure}
Training was performed using the loss function in Eq.~(\ref{eq_loss_wq}).
Please refer to the supplemental material for details on the parameters used in the training.
The estimation results for the constrained and unconstrained cases are shown in Figs.~\ref{fig4}(a) and \ref{fig4}(b).
Also, please refer to Appendix~\ref{sec_video}, which contains video information regarding the estimation results of the Hamiltonian time series. 
As shown in the results, the introduction of symmetry constraints allowed the neural network modeled Hamiltonian function (heat map) to capture the features of the original Hamiltonian function (red contour lines) set at the time of dataset generation.
In the case of without constraints, a Hamiltonian significantly deviating from the original Hamiltonian was learned. 
This result was confirmed not only through visual comparison but also through quantitative comparison of the similarity between the true Hamiltonian and the Hamiltonian estimated by the DNN [Table~\ref{tbl1}].
From the values of the loss function for the validation data revealed that the unconstrained case had better prediction performance than the constrained case [Table~\ref{tbl1}].
This comparison corresponds to the search for the hyperparameter $\lambda$ associated with {\bf step 3} .
The reversal of the Hamiltonian estimation accuracy and prediction performance indicates the danger of determining the hyperparameters based on the prediction performance, as described in {\bf step 3}. 
\begin{table}[b]
\centering
\caption{Cosine similarity between the Hamiltonian function estimated by DNN and the Hamiltonian function set when generating the data, 
and mean value of the first term of the loss function (the first term of Eq.~(\ref{eq_loss_wq})) for the validation data. 
Mean $\pm$ standard error of Cosine similarity and prediction errors over 30 independent trials.
Note that higher is better for cosine similarity; lower is better for L2-loss.}
\label{tbl1}
\begin{tabular}{lcc}
\toprule
\textbf{Method} & \textbf{cosine similarity} & \textbf{L2-loss} \\
\midrule
W/ constraint  & {\bf 0.45} $\pm$ 0.29 & (5.46 $\pm$ 0.48) $\times$ $10^{-8}$  \\
W/O constraint  & 0.06 $\pm$ 0.09 & ({\bf 2.58}  $\pm$ 0.27) $\times$ $10^{-8}$  \\
\bottomrule
\end{tabular}
\end{table}

\section{Summary and Discussion}
\label{sec_summary}
In this paper, we propose a framework for evaluating the structural uncertainty arising in physics-informed machine learning for physical model estimation. 
The proposed method was verified in both simple systems, such as spring motion, and more complex systems, such as wave motion equations, confirming its effectiveness. \par
As stated in Introduction, there are three types of uncertainty: structural uncertainty, model-form uncertainty, and data uncertainty. 
We proposed a method for evaluating the structural uncertainty. 
On the other hand, in actual demonstrations, data is finite, so the effect of the data uncertainty could potentially arise.
As shown in Appendix~\ref{sec_appendixB}, when estimating the derivative of the coefficient function without using a neural network, it was observed that the finiteness of the data significantly affected the estimation results. 
This difference is considered to arise because, in this paper, we estimated the coefficient function itself using PIML~(see Appendix~\ref{sec_appendixB}). 
Thus, it was confirmed that using PIML partially mitigates the data uncertainty. 
\par
The Limitation of this study is that the theoretical foundation presented in this paper is built under idealized assumptions, namely, that the PDE is linear with respect to the coefficient function. 
We consider that there is considerable room for further theoretical development beyond these assumptions.
In particular, there should be a possible theoretical extension of uncertainty analysis for nonlinear PDEs, in which the coefficient function enters nonlinearly. 
For the first direction, we believe that singular learning theory \citep{watanabe2009algebraic} offers a promising approach. 
This theory can evaluate the non-uniform loss landscape, and that is why it has been applied to quantify uncertainties in deep learning models, and recent studies have started leveraging it for uncertainty evaluation in model selection and generalization \cite{wang2024loss}. 
However, to the best of our knowledge, no existing studies have succeeded in evaluating the degree and structure of uncertainty quantitatively, particularly for each parameter. 
We believe this direction offers significant potential.\par
As we mentioned, the proposed method is limited to PDEs with linear partial differential coefficients; however, many practical physical models belong to this class. That is why the proposed method is expected to have broader impacts on a wide range of future scientific research using machine learning.




\bibliography{main}
\bibliographystyle{tmlr}

\appendix
\section{Appendix: DNN Model and its Training Parameters}
\label{sec_dnn_params}
Here, we describe the DNN models and their training settings. 
In this study, we used a fully coupled multilayer neural network as the DNN model. 
The DNNs consisted of an input layer, two hidden layers, and an output layer. 
The number of nodes in each layer was set as shown in the ``Network structure'' in Table~\ref{tbl_dnnsettings}. 
The activation functions of the DNNs were set as the hyperbolic tangent function as shown in the ``Activation function'' in Table~\ref{tbl_dnnsettings}. 
The tanh function is defined as
\begin{equation}
\tanh(x) = \frac{\exp(x) - \exp(-x)}{\exp(x) + \exp(-x)}.
\end{equation}\par
The number of samples used for training DNN is shown in Table~\ref{tbl_dnnsettings} as ``Training data size $N$''. 
The Adam method~\citep{kingma2014adam} was used for training. 
The training iterations are shown in Table~\ref{tbl_dnnsettings}. 
For other details, please see the code shared as follows:
\url{https://anonymous.4open.science/r/Structural_uncertainty-30D5}
\begin{table}[h]
\centering
\caption{Parameters of DNN model and its training. In the ``Network structure'', the number of nodes is shown in the order from left to right: input layer -- first layer -- second layer -- third layer -- output layer.}
\label{tbl_dnnsettings}
\begin{tabular}{|c|c||c|c|}\hline
\textrm{Parameter name}& & \textrm{Parameter name} & 
\\ \hline
Training data size $N$ & 10,000 & Network structure & 2-100-10-1 \\
Activation function & tanh & Training iteration& 400,000\\
Training algorithm & Adam & & \\ \hline
\end{tabular}
\end{table}

\section{Bridging the Assumption of Infinitesimal Grids in Uncertainty Evaluation and Learning with Finite Data using DNN Model}
\label{sec_appendixB}
\begin{figure}[t]
\centering
\includegraphics[width=0.5\linewidth]{"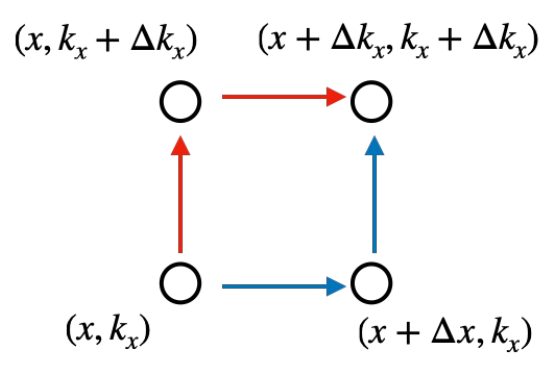"}
\caption{Conceptual diagram of different integration paths in $(x,k_x)$ space to obtain the coefficient function. The red path to integrate $k_x$ first and the blue path to integrate $x$ are shown.}
\label{sup_fig1}
\end{figure}

\begin{figure}[bh]
\centering
\includegraphics[width=\linewidth]{"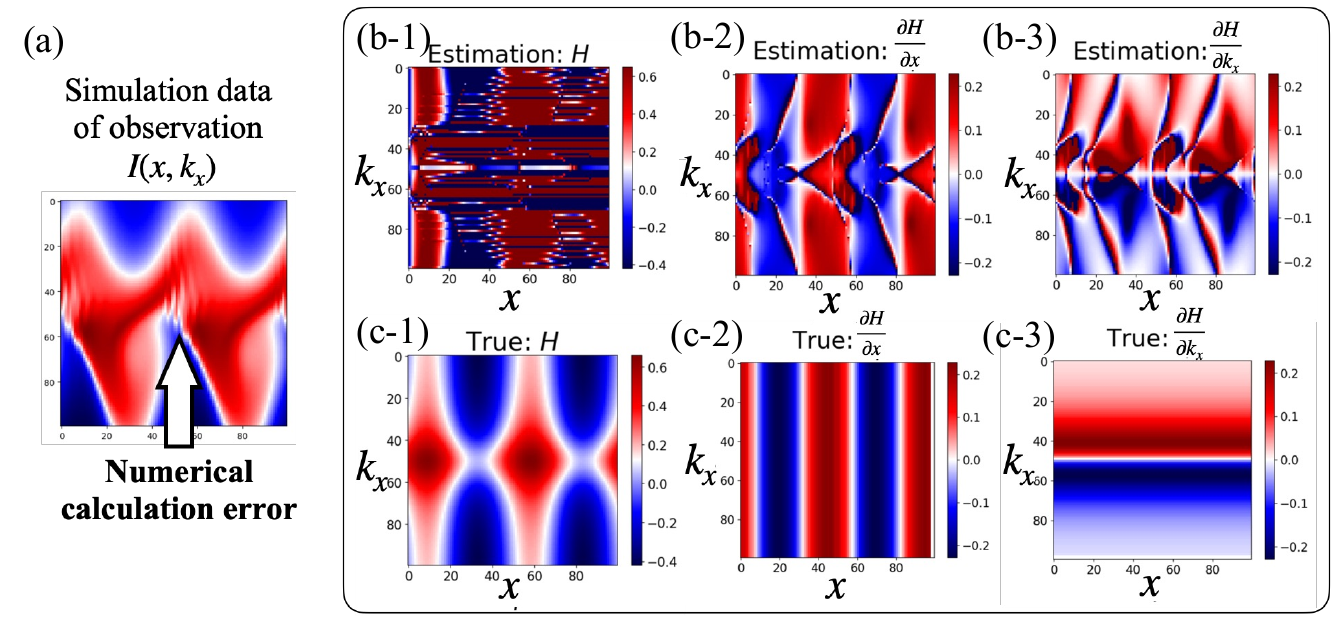"}
\caption{Hamiltonian estimation results from the inverse matrix computation of $M$.
(a) Simulated observational data $I(x,k_x)$ used for estimation. The observational data have numerical errors. (b) Estimation results of the Hamiltonian function and its partial differential coefficients. (c) True Hamiltonian and its partial differential coefficients.}
\label{sup_fig2}
\end{figure}

As we noted in ``Sec.~\ref{sec_summary} Summary and Discussion'', there are significant differences between the evaluation of uncertainties and the estimation of the coefficient function by machine learning. 
The difference is that uncertainties were evaluated on the basis of the assumption of an infinite number of data points on an infinitesimal grid, whereas finite data was used in the estimation of coefficient functions by machine learning. 
This difference is critical. 
For example, even if the partial differential coefficients of all Hamiltonians were known on a finite grid, different Hamiltonians would be estimated for different integral paths. 
For example, if there are red and blue paths as shown in Fig.~\ref{sup_fig1}, the integral at $(x+\Delta x,k_x + \Delta k_x)$ may change depending on which path is taken. 
The reason is that the constraint on the consistency of the partial differential coefficients due to the different integration paths is not included when estimating the partial differential coefficients. 
For example, this inconsistency is eliminated by the following constraints: 
\begin{eqnarray}
\label{eq_consist}
\left(\left.\frac{\partial H}{\partial x}\right|_{(x,k_x)}+\left.\frac{\partial H}{\partial k_x}\right|_{(x+\Delta x,k_x)}\right) - \left(\left.\frac{\partial H}{\partial k_x}\right|_{(x,k_x)}+\left.\frac{\partial H}{\partial x}\right|_{(x,k_x +\Delta k_x)}\right)=0.
\end{eqnarray}
Furthermore, although the partial differential coefficients were assumed to be independently determined during the uncertainty evaluation, the actual coefficient functions are likely to be continuous functions, so the partial differential coefficients in the neighborhood will be correlated. 
These constraints also need to be introduced into the machine learning algorithm. 
Moreover, although it was assumed that no noise was added to the data during the uncertainty evaluation, noise is added to real data. 
For example, in the simulation data for the observational data $I(x,k_x)$ used in this study, numerical calculation errors are added to the data (Fig.~\ref{sup_fig2}).\par
Substituting the observational data set $I(x,k_x)$ into the following equation used to evaluate uncertainty, we can estimate the Hamiltonian function by inverse matrix $M$ calculation.
\begin{eqnarray}
\left(\begin{matrix}
I_t(x^{(1)},k_x^{(1)}) - C(x^{(1)},k_x^{(1)})\\
I_t(x^{(1)},-k_x^{(1)}) - C(x^{(1)},-k_x^{(1)})\\
\vdots\\
I_t(x^{(N_x)},k_x^{(N_{k_x}/2)}) - C(x^{(N_x)},k_x^{(N_{k_x}/2)})\\
I_t(x^{(N_x)},-k_x^{(N_{k_x}/2)}) - C(x^{(N_x)},-k_x^{(N_{k_x}/2)})
\end{matrix}\right)
=\boldsymbol{M}
\left(\begin{matrix}
H_x(x^{(1)},k_x^{(1)})\\
H_{k_x}(x^{(1)},k_x^{(1)})\\
\vdots\\
H_x(x^{(N_x/2)},k_x^{(N_{k_x}/2)})\\
H_{k_x}(x^{(N_x/2)},k_x^{(N_{k_x}/2)})
\end{matrix}\right),
\end{eqnarray}
\footnotesize
\begin{eqnarray}
\boldsymbol{M}=
\left(\begin{matrix}
-I_{k_x}(x^{(1)},k_x^{(1)})&I_x(x^{(1)},k_x^{(1)})& & & & & \\
-I_{k_x}(x^{(1)},-k_x^{(2)})&I_x(x^{(1)},-k_x^{(2)})& & &\text{\huge{0}} & &\\
 & & &\ddots & \\
  & \text{\huge{0}}& & &-I_{k_x}(x^{(N_x/2)},k_x^{(N_{k_x}/2)})&I_x(x^{(N_x/2)},k_x^{(N_{k_x}/2)})&\\
  & & & &-I_{k_x}(x^{(N_x/2)},-k_x^{(N_{k_x}/2)})&I_x(x^{(N_x/2)},-k_x^{(N_{k_x}/2)})&\\
\end{matrix}\right),\nonumber
\end{eqnarray}
\normalsize
where the matrix size of $\mathbf{M}$ is $N_xN_{k_x} \times N_xN_{k_x}$. 
If $M$ is full rank and the data points are given in an infinitesimal grid, it should be possible to estimate the coefficient function, i.e. the Hamiltonian function, in this way as well. 
In fact, the matrix $M$ was numerically full rank. 
However, the results of estimating their partial differential coefficients were disastrous. 
$\frac{\partial H}{\partial x}$ and $\frac{\partial H}{\partial k_x}$ have structures far from the true $\frac{\partial H}{\partial x}$ and $\frac{\partial H}{\partial k_x}$ at around $k_x=0$ [Figs.~\ref{sup_fig2}~(b-2), \ref{sup_fig2}~(b-3), \ref{sup_fig2}~(c-2), and \ref{sup_fig2}~(c-3)]. 
As a result, the estimation of the Hamiltonian function was also very inaccurate [ Figs~\ref{sup_fig2} (b-1) and (c-1)]. 
The results of the Hamiltonian estimation varied considerably depending on the path of numerical integration used to estimate the Hamiltonian function. 
The cause of these worse estimation results arises from the numerical error added to the observational data $I(x,k_x)$ and the inconsistency of the partial differential coefficients due to the integration path.\par
In the PIML approach, since the coefficient function itself is estimated, the inconsistency problem arising from the integration path mentioned above when estimating the derivative of the coefficient function does not occur. 
Thus, the use of DNNs is expected to mitigate to some extent the discrepancies from the indefinite evaluation time due to the finiteness of the data, and the accurate estimation results of the Hamiltonian function presented in this study guarantee that this is the case.\par

\section{Video of Hamiltonian Estimation Results}
\label{sec_video}
Please refer to the attached files of ``movie.gif'', with symmetry constraints, and ``movie\_withoutconst.gif'', without symmetry constraints, at \url{https://anonymous.4open.science/r/Structural_uncertainty-30D5}. 
The files show all the estimated Hamiltonian functions at each time as movies.

\end{document}